\font\Bbb=msbm10
\let\mcd=\mathchardef
\mcd\Pe="7C50
\def\12{{\textstyle{1\over2}}}
\documentclass[12pt]{amsart}
\usepackage{graphics}
\usepackage{xcolor}
\begin{document}
\title{Is superhydrophobicity robust with respect to disorder?}
\author{Jo\"el De Coninck}
\address{Laboratory of Surface and Interfacial Physics, University of Mons,
Avenue Maistriau, 19, 7000 Mons, Belgium} 
\email{joel.deconinck@umh.ac.be}
\author{Fran\c cois Dunlop}
\author{Thierry Huillet}
\address{Laboratory of Physics: Theory and Models,
CNRS UMR 8089, Cergy-Pontoise University, 95302 Cergy-Pontoise, France}
\email{francois.dunlop@u-cergy.fr}
\email{thierry.huillet@u-cergy.fr}

\abstract{
We consider  theoretically
the Cassie-Baxter and Wenzel states describing the wetting
contact angles for rough substrates. More precisely, we consider
different types of periodic geometries such as square protrusions and
disks in 2D, grooves and nanoparticles in 3D and derive explicitly the
contact angle formulas. We also show how to introduce the concept of
surface disorder within the problem and, inspired by biomimetism,
study its effect on superhydrophobicity. Our results, quite generally,
prove that introducing disorder, at fixed given roughness,
will lower the contact angle: a disordered substrate will have a lower contact 
angle than a corresponding periodic substrate. We also show that
there are some choices of disorder for which the loss of
superhydrophobicity can be made small, making superhydrophobicity robust. \\

\noindent
{\sl PACS:}
{68.08.Bc} {Wetting} - 
{68.35.Ct} {Interface structure and roughness} - 
{05.70.Np} {Interface and surface thermodynamics}
}
}
\maketitle
\section{Introduction}
\begin{figure}
 \includegraphics{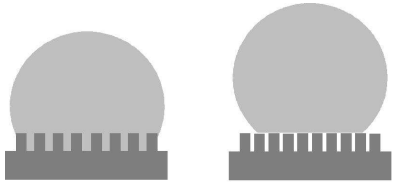}
\caption{\label{fig1}Wenzel state (left), Cassie-Baxter state (right)}
 \end{figure}

Superhydrophobic surfaces have attracted  tremendous interest in the
last few years, not only for academic reasons
\cite{RSN,ZSNJW,GE,Q,YTP,S,B,RVVSCBID,DGDRVSVDJJDL,GVDPJDJ}. 
Nowadays, everybody agrees about the phenomenological definition of
such surfaces. When a water droplet is deposited, high values of both
advancing and receding static contact angles are
observed. Superhydrophobicity also  induces special characteristics
such as the rolling  of a deposited water drop at a very low tilt
angle (equivalently, a very small hysteresis) or the rebound of a drop
on impact with the surface. These properties are often described as
self-cleaning or the ‘Lotus’ effect.  
In any case the presence of a certain kind of roughness, which is
still to be defined precisely, is necessary. On top of such surfaces,
it is expected that the drop can be in at least two different states:
in contact everywhere with the solid surface, i.e. the so-called wet
or Wenzel state, or in contact with only the top elements of the
surface, the so-called  dry or Cassie-Baxter state. 
Superhydrophobicity refers naturally to this Cassie-Baxter state. 
A schematic representation of both states is given in Fig. \ref{fig1}. 
If we compare the free energies of both situations 
for a periodic substrate, we arrive at the
classical conclusion that if the equilibrium contact angle $\theta_0$
corresponding to the flat surface satisfies some inequality, the drop
will be in the Cassie-Baxter state leading to superhydrophobicity. 
This inequality is well known and can be written as 
\begin{equation}\label{1} 
\cos\theta_0 < -(1-\phi)/(r-\phi)
\end{equation} 
where $r$ is the Wenzel roughness of the surface, defined as the total
length (area, in 3D) of the surface divided by its projection, and
$\phi$ is the covered fraction, defined as the total length, or area
in 3D, at contact divided by the total projected length. As can be
easily seen, the larger $r$, the more the surface is likely to be in the
Cassie-Baxter state. This is the reason why it is believed that a very
rough hydrophobic surface can be superhydrophobic. 

The classical argument sketched above neglects meta\-stability,
as will be the case throughout the present paper.
Having in mind low
viscosity liquids such as water, we focus on the minimum free energy
configuration without taking into account hysteresis effects where
dynamics has to be considered. This approach also excludes large defects, wells 
or spikes or chemical impurities of size say $100\,\mu$m or bigger,
which may be present by fabrication, by nature, or by large deviations of 
disorder.
We assume a typical defect size say $10\,\mu$m or smaller  
and drop size or contact line length much bigger than the typical distance 
between defects.
A self-averaging hypothesis can then be used in the thermodynamic limit.
For a discussion of metastability associated with a large single defect,
in models such as discussed in the present paper, see \cite{DDH11}.
For a discussion of metastability associated with large deviations of disorder,
see \cite{CDDR,CDDD}.

From an experimental point of view, it has been noticed that
composite superhydrophobic states can be observed on surfaces with
Wenzel roughness $r$ as low as 1.25 \cite{RDDVD}.  It is also clear
that reducing the details of the topography of the surface $z(x,y)$ to
a unique parameter such as the Wenzel roughness $r$, as in (\ref{1}),
is a very crude approximation.  Recently, an  index has been introduced to
characterize for a given surface, which part of the surface is
superhydrophobic and which part  is not.  
This so-called index of superhydrophobicity \cite{RDDVD} allows to see
which part of the surface is in a Cassie-Baxter state and which part
is in a Wenzel state. 
Unfortunately, it does not explain how the details of the roughness
can affect the robustness of superhydrophobicity.
Will a small change in roughness induce a big change in superhydrophobicity?

There are in fact two ways to design superhydrophobic surfaces: with a
periodic topography using some synthetic procedure for very small
areas by lithography or deep reaction ion etching techniques or with
some disorder using a natural self-organising process for large
areas. Of course, the fabrication of these highly precise
microstructures is very complex. The only hope to have a scalable
procedure in practice is  thus dependent on
self-organisation. Moreover, the first  types of surfaces are
typically based on single-scale topography and the second sets,
usually, present several scales as for example do leaves \cite{BN}. This has
already been observed experimentally. For instance, when several
layers of nanoparticles are deposited on a glass substrate by
Langmuir-Blodgett techniques, it is easily seen that the distribution
of particles in the first layer is different from the one in the
second layer. Therefore, we should observe  different behavior
depending on the number of layers. In fact, the disorder will increase
with the number of layers. For such kinds of rough substrate, it has
been shown experimentally \cite{TYL} that disorder leads to a
reduction of superhydrophobicity.  How much hydrophobicity is lost is
not yet clear so far. In Nature, there are many examples of plants or
animals that have this remarkable property of superhydrophobicity
without exactly periodic surface topography. Topographical defects
observed in the living world do not seem to affect the robustness of
the superhydrophobicity. 
It is precisely the aim of the present paper to introduce disorder
within superhydrophobicity and to analyze how it will
affect the Lotus effect. 

\section{Square protrusions}
\begin{figure}
\resizebox{0.48\textwidth}{!}{\includegraphics{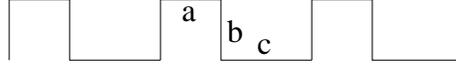}}
\caption{\label{fig2}The periodic surface characterized by three
  parameters: $a$, $b$ and $c$.} 
\end{figure}

\begin{figure}
\resizebox{0.78\textwidth}{!}{\includegraphics{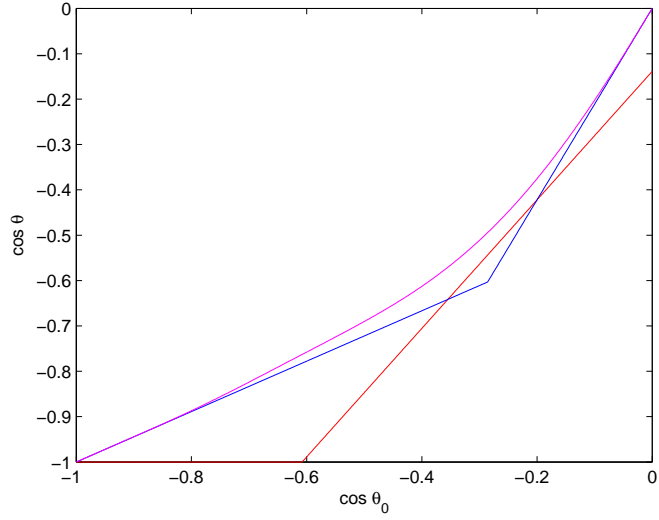}}
\caption{\label{fig3}Square protrusions: $\cos\theta$ versus $\cos\theta_0$, 
with $a/b=1$, $\langle c\rangle/b=0.8$. From bottom-right to top-left:
Cassie-Baxter and Wenzel, periodic (blue), Bernoulli distribution
with $c_1/b=0.5,\,c_2/b=1.1$ (central part of red curve), exponential
distribution, eq. (\ref{thetaexp}) (magenta).}
\end{figure}

Let us start with a simple example. 
In two dimensions, or in three dimensions with grooves, we can build a
regular surface such as the one illustrated in Fig. \ref{fig2}. 
Denoting $\gamma_{SL}$, $\gamma_{SV}$, $\gamma_{LV}$, the flat solid/liquid,
solid/vapor, liquid/vapor surface tensions, and using superscripts
$CB$ for Cassie-Baxter and $W$ for Wenzel, we have free energies for one motif, 
\begin{eqnarray}\label{CBW}
(a+c)\gamma_{SL}^{CB} &=&a\gamma_{SL}+2b\gamma_{SV}+c(\gamma_{LV}+\gamma_{SV} )\cr
(a+c)\gamma_{SL}^W &=& (2b+c+a)\gamma_{SL}
\end{eqnarray}
Using Young's equation for the flat substrate, where the contact angle is
$\theta_0$,  we obtain that the lower energy state is Cassie-Baxter (state 1) if
$c<c_{12}(\theta_0)$, where
\begin{equation}
c_{12}(\theta_0)=-2b\cos\theta_0/(1+\cos\theta_0)
\end{equation}
 and Wenzel (state 2) if
$c>c_{12}(\theta_0)$, with coexistence at $c=c_{12}(\theta_0)$.  
The Wenzel roughness of the motif is $r(c)=1+2b/(a+c)$ and the covered
fraction is $\phi(c)=a/(a+c)$. Then $c<c_{12}(\theta_0)$ is equivalent
to inequality (\ref{1}). In the periodic case, the
lengths $a$, $b$ and $c$ are the same everywhere, so that either
Cassie-Baxter or Wenzel will be the winner everywhere.
The resulting contact angle on the periodic substrate is $\theta=\theta^{CB}$ 
obtained with $\gamma_{SL}^{CB}$ when $c<c_{12}(\theta_0)$ and $\theta=\theta^W$ 
obtained with $\gamma_{SL}^W$ when $c>c_{12}(\theta_0)$, as shown on 
Fig.~\ref{fig3}, with  
\begin{eqnarray}\label{thetapro}
\cos\theta^{CB}&=&\phi\cos\theta_0+\phi-1\cr
\cos\theta^W&=&r\cos\theta_0
\end{eqnarray}
If $\theta_0>3\pi/4$,
not a realistic case, a state with water reaching the bottom but
keeping air in the corners may be the winner. We ignore it here for
simplicity, but air trapped at the bottom corners will be included
below when considering 2D-disks or 3D-nanoparticles. 

Now, let us assume that the square protrusions are distributed with
some disorder, in the sense that $c$ takes 
values $c_1,c_2,c_3,\dots$ distributed according to some stationary
probability law. What will be the effect on superhydrophobicity?
Is a natural surface less or more superhydrophobic than a synthetic one?  
Clearly, we will have a non-trivial combination of Wenzel and
Cassie-Baxter states depending on the probability distribution of
$c$. For a given sequence $c_1,c_2,\dots,c_n$ we get a free energy,
defining a composite solid-liquid surface tension  $\gamma_{SL}^{rr}$,
with a superscript $rr$ for the random roughness,
\begin{eqnarray*}
(na+\sum_ic_i)\gamma_{SL}^{rr}=\big(na+2b(n-n_1)+\sum_{i:W}c_i\big)\gamma_{SL}+\cr
+2bn_1\gamma_{SV}+\sum_{i:CB}c_i(\gamma_{LV}+\gamma_{SV})
\end{eqnarray*}
where $\sum_{i:CB}$ refers to the Cassie-Baxter states over the length $n$,
the number of which equals $n_1$, and $\sum_{i:W}$ refers to the
Wenzel states over the length $n$, the number of which equals
$n_2=n-n_1$. We then get in the thermodynamic limit $n\to\infty$,
assuming that the contact line samples a large number $n$ of protrusions,
\begin{eqnarray}\label{gaslpro}
(a+\langle c\rangle)\gamma_{SL}^{rr}=
a\gamma_{SL}+2bp_1\gamma_{SV}+p_1c_1(\gamma_{LV}+\gamma_{SV})+\cr
+(2bp_2+p_2c_2)\gamma_{SL}\hskip1cm
\end{eqnarray}
where $p_1$ is the probability (or limiting fraction) that $c<c_{12}(\theta_0)$,
and $p_2=1- p_1$, and $c_1$ is the average of $c$ conditioned on
$c<c_{12}(\theta_0)$, and similarly $c_2$, so that 
$\langle c\rangle=p_1c_1+p_2c_2$. 
Note that $p_{1}$ (respectively 
$p_{2}$) is as well the probability to be in a Cassie-Baxter state
(respectively Wenzel), emphasizing that in the randomized description of the
wetting problem, the pure states no longer exist. 
The contact angle $\theta$ is given by Young's equation as
\begin{equation}\label{thetaabc} 
\cos\theta= {\gamma_{SV}^{rr}-\gamma_{SL}^{rr}\over\gamma_{LV}}
={(a+2bp_2+p_2c_2)\cos\theta_0-p_1c_1\over a+\langle c\rangle} 
\end{equation} 
where $\gamma_{SV}^{rr}$ is the composite solid-vapor surface tension, 
$$
\gamma_{SV}^{rr} = \gamma_{SV} {a+2b+\langle c\rangle\over a+\langle c\rangle}
$$
An interesting property is that $d \cos\theta/ d \cos\theta_0$ is
always equal to the roughness $r$ when $\cos \theta_0=0$ and to the
fraction $\phi$ when $\cos \theta_0=-1$. In between, $\cos\theta$ is not in
general a linear function of $\cos \theta_0$ because $p_1,p_2,c_1,c_2$
depend upon $\theta_0$, but it is convex: when the random variable $c$ has a
density $f(c)$ with respect to the Lebesgue measure, a simple computation gives
\begin{equation}
{d^2\cos\theta\over d(\cos\theta_0)^2}={4b^2\over a+\langle c\rangle}
{f(c_{12})\over(1+\cos\theta_0)^3}\ge0
\end{equation}
Concerning superhydrophobicity, we get ($c_{12}:=c_{12}\left(\theta_{0}\right)$)
\begin{eqnarray}
\cos\theta-\cos\theta^{CB}&=& 
p_2(1+\cos\theta_0){c_2-c_{12}\over a+\langle c\rangle}\label{CB}\\
\cos\theta-\cos\theta^W&=&
p_1(1+\cos\theta_0){c_{12}-c_1\over a+\langle c\rangle} \label{W}
\end{eqnarray}
where $\cos\theta^{CB}$ and $\cos\theta^W$  are obtained from
(\ref{CBW}) with $c$ replaced by $\langle c\rangle$, using Young's
equation, as if only  Cassie-Baxter configurations or only Wenzel configurations
were present. Both (\ref{CB}) and (\ref{W}) and positive. This shows that any
disorder will lower 
the contact angle $\theta$. Naturally, there will always be some  specific
distribution that minimizes this difference, which could be
significant for practical applications. 

These expressions  can be evaluated for any probability distribution for the
$c$'s. The resulting plots of $\cos\theta$ versus $\cos\theta_0$ are shown 
in Fig.~\ref{fig3} for two examples discussed below, together with the periodic 
case.

As a first example, the squares may be distributed randomly
uniformly on the surface, with the $c$'s exponentially distributed, leading to 
\begin{equation}\label{thetaexp} 
\cos\theta=\phi\cos\theta_0+(1-\phi)\big((1+\cos\theta_0)
e^{{r-1\over1-\phi}{\cos\theta_0\over1+\cos\theta_0}} -1\big)
\end{equation} 
which shows interestingly that the Wenzel roughness $r$  is not enough
to describe the properties of the surface.  
Such disorder should be expected in the case of adsorption from a gas phase or a
solvent.

A second example is a Bernoulli variable taking value $c_1$ with probability
$p_1$ and value $c_2$ with probability $p_2$, with parameters such that
$c_1<c_{12}<c_2$. Then (\ref{thetaabc}) holds with $p_1,p_2,c_1,c_2$ now 
independent of $\theta_0$ and here $\cos\theta$ is linear in $\cos \theta_0$ 
in the corresponding range. The associated result corresponds to
the central part of the red line in Fig.~\ref{fig3} supplemented by a piece of
the Cassie-Baxter straight line on the left and a piece of the Wenzel straight
line on the right.

\section{Disks}
\begin{figure}
\resizebox{0.48\textwidth}{!}{\includegraphics{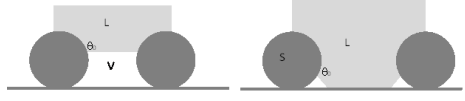}}
 \caption{\label{fig4}The Cassie-Baxter state (left) and Wenzel state (right)
with disks. The contact angle of the liquid with the solid surface of
   the disk and with the solid at the bottom is $\theta_0$.} 
 \end{figure}

\begin{figure}
\resizebox{0.78\textwidth}{!}{
\includegraphics{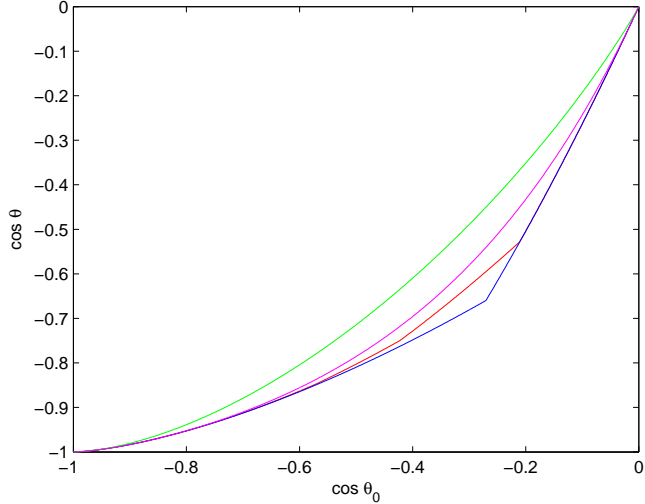}
}
\caption{
Disks: $\cos\theta$ versus $\cos\theta_0$ with
$\langle c\rangle/D=0.8$. From bottom-right to top-left:
Cassie-Baxter and Wenzel, periodic (blue), $d$-geometric model with $d/D=0.8$
(red), exponential model (magenta), Gamma model with $\alpha=0.2$ (green).
} 
\label{fig5}
\end{figure}

Let us now consider the case of disks with diameter $D$ distributed
along a line with distance $c$ between successive disks. The disks are made of
the same material as the plane substrate. The dry and
wet cases are presented in Fig. \ref{fig4}. 
The Wenzel state only exists when $c>c_{\min}$, where
\begin{equation}
c_{\min}=-D(1+2\cot\theta_0)
\end{equation}
When $c<c_{\min}$, the two traps of air shown on Fig~\ref{fig4} in the Wenzel 
case would be overlapping. The condition $c>c_{\min}$
is a restriction only when $c_{\min} >0$ or 
$\theta_0>\arctan(-2)\simeq116.6^\circ$. However, it can be checked that, for
all $\theta_0$ in $(\pi/2,\pi)$: $c_{12}>c_{\min}$. Thus,
consistently, Wenzel states characterized by $c>c_{12}$ always obey
$c>c_{\min}$.  The same reasoning as for the squares leads in the
periodic case to
\begin{eqnarray}\label{CBWdisks}
(a+c)\gamma_{SL}^{CB} &=&A_{1}\gamma _{SL}+\left( B_{1}+c\right) \gamma
_{SV}+\left( C_{1}+c\right) \gamma _{LV}\cr
(a+c)\gamma_{SL}^W &=&\left( A_{2}+c\right) \gamma _{SL}+B_{2}\gamma
_{SV}+C_{2}\gamma _{LV}
\end{eqnarray}
with
$$
A_{1}=D\left( \pi -\theta _{0}\right),\ B_{1}=D\left( 1+\theta
_{0}\right),\ C_{1}=D\left( 1-\sin \theta _{0}\right)
$$
and
\begin{eqnarray*}
A_{2}=D\left( 1-2\theta _{0}+2\cot \theta _{0}+2\pi \right)\hskip3cm\cr
B_{2}=D\left( 2\theta _{0}-2\cot \theta _{0}-\pi \right),\quad
C_{2}=D\frac{%
1+\cos \left( 2\theta _{0}\right) }{\sin \theta _{0}}
\end{eqnarray*}
The resulting Cassie-Baxter and Wenzel contact angles on a periodic arrangement 
of disks in 2D or tubes in 3D are 
\begin{eqnarray}\label{thetanano}
\cos\theta^{CB}&=&{D(\pi-\theta_0)\cos\theta_0-D(1-\sin\theta_0)-c\over D+c}\cr
\cos\theta^W&=&{(D(2\pi+1-2\theta_0)+c)\cos\theta_0\over D+c}
\end{eqnarray}
From (\ref{CBWdisks}) we have $\gamma_{SL}^{CB}<\gamma_{SL}^W$ when 
$c<c_{12}(\theta_0)$, and conversely, with
\begin{equation}\label{c12disks}
{c_{12}(\theta_0)\over D}=
{\cos\theta_0(\theta_0-\pi-1)-1+\sin\theta_0\over1+\cos\theta_0}
\end{equation}
Therefore, the true contact angle on such a periodic substrate is 
$\theta=\theta^{CB}$ when $c<c_{12}(\theta_0)$ and $\theta=\theta^W$ 
when $c>c_{12}(\theta_0)$. A corresponding plot of $\cos\theta$ versus 
$\cos\theta_0$ is shown in Fig.~\ref{fig5}.

In the random case, with $p_1,p_2,c_1,c_2$ defined as after (\ref{gaslpro}),
using $c_{12}(\theta_0)$ in (\ref{c12disks}), one may compute
$$
\gamma _{SV}^{rr}=\gamma _{SV}\,\frac{\pi D+D+\langle c\rangle}
{D+\left\langle c\right\rangle}
$$
and
\begin{eqnarray*}
(D+\langle c\rangle)\gamma _{SL}^{rr}
=p_{1}( A_{1}\gamma _{SL}+B_{1}\gamma _{SV}+C_{1}\gamma_{LV}) \cr
+p_{2}(A_{2}\gamma _{SL}+B_{2}\gamma _{SV}+C_{2}\gamma _{LV}) \cr
+p_1c_1( \gamma _{LV}+\gamma _{SV}) + p_2c_{2}\gamma_{SL}
\end{eqnarray*}
leading to
\begin{eqnarray}\label{nanoCB}
\cos\theta-\cos\theta^{CB}
&=& p_2(1+\cos\theta_0){c_2- c_{12}\over D+\langle c\rangle}\label{CBdisk} \\
\cos\theta-\cos\theta^W
&=& p_1(1+\cos\theta_0){c_{12}-c_1\over D+\langle c\rangle}    \label{Wdisk}
\end{eqnarray}
where $\cos\theta^{CB}$ and $\cos\theta^W$  are given by
(\ref{thetanano}) with $c$ replaced by $\langle c\rangle$, 
which is very similar to (\ref{CB})(\ref{W}). Again (\ref{CBdisk}) and
(\ref{Wdisk}) are positive.

The distribution of disks might be considered to be like a
quenched hard sphere fluid. In one dimension the distance $c\ge0$
between neighboring disks would then follow an exponential
distribution, with
$$
p_1=1-e^{-c_{12}/\left\langle c\right\rangle }\,,\quad p_1c_{1}=\left\langle
c\right\rangle -\bigl( \left\langle c\right\rangle +c_{12}\bigr)
e^{-c_{12}/\left\langle c\right\rangle } 
$$ 
But disks may have a small attractive
interaction, electrostatic or due to the solvent, before drying. Then
the distribution of inter-particle distances will be more weighted at
small distances, at the expense of  distances around the range of the
interaction. This may be modeled for instance by a Gamma distribution,
where the probability density of $c$ is 
$$
f(c)={\beta^\alpha\over\Gamma(\alpha)}\,c^{\alpha-1}e^{-\beta c}
$$
with $\beta=\alpha/\langle c\rangle$, from which (\ref{CBdisk})(\ref{Wdisk}) can
be computed with
$$
p_1=1-\int_0^{c_{12}}f(c)dc,\quad p_1c_1=\int_0^{c_{12}}cf(c)dc 
$$
In Fig. \ref{fig5} we choose $\alpha=0.2$, corresponding to more $c$'s both at 0
and at infinity compared to the exponential distribution with the same
$\langle c\rangle$. The Gamma model is the highest curve on
Fig. \ref{fig5}, and next is the curve for the exponential
model. Globally, superhydrophobicity is reduced. In the limit 
$\alpha,\,\beta\searrow0$ keeping $\alpha/\beta=\langle c\rangle$ fixed,
producing large clusters of disks, superhydrophobicity is completely lost:
$\cos\theta=\cos\theta_0$. This is the worst case given the convexity property.

In the opposite direction, disks could be deposited on a
structured substrate, where $c$ would take values on a discrete set
corresponding to the minima of the structured substrate. For example,
on top of a compact arrangement of disks of diameter $d$,
neighboring disks of diameter $D$ bigger than $d$ but smaller
than $2d$ would be at distances $c$ obeying $c+D$ in the range 
$\{2d, 3d,\dots\}$. 
Experimentally, this could be achieved for instance by Langmuir-Blodgett
depositions or successive spin coatings.
On this set we choose a geometric distribution, with the
same $\langle c\rangle$ as the other models: 
$$
\Pe(c=nd)={1-e^{-\lambda d}\over e^{-2\lambda d}}e^{-\lambda nd}\,,\qquad n\ge2
$$
with
$$
\lambda={1\over d}\log{D+\langle c\rangle-d\over D+\langle c\rangle-2d} 
$$
from which (\ref{CBdisk})(\ref{Wdisk}) can be computed with
$$
p_2=\exp\Bigl(-\lambda d\Bigl\lceil{c_{12}+D\over d}\Bigr\rceil\Bigr)
$$
$$
p_2c_2={p_2d\over1-e^{-\lambda d}}\Bigl(\Bigl\lceil{c_{12}+D\over d}\Bigr\rceil
(1-e^{-\lambda d}\Bigr)+e^{-\lambda d}\Bigr)-Dp_2
$$
where $\lceil x\rceil$ is the smallest integer above $x$.
One obtains for $\cos\theta$ as function of $\cos\theta_0$ a
continuous broken curve with infinitely many discontinuities of the
derivatives accumulating at $\cos\theta_0=-1$. 
A discontinuity occurs at every value of $\cos\theta_0$ such that
$c_{12}(\theta_0)+D$ is a multiple integer of $d$.
As Fig. \ref{fig5} shows, this $d$-geometric model really enhances
superhydrophobicity compared to the exponential model. 

The analysis also allows to compare the effect of square blocks with
that of disks on superhydrophobicity. We find that disks are more favorable
since they increase a little more the contact angle $\theta$: 
compared to Fig. \ref{fig3}, the curves on Fig. \ref{fig5}, starting with
the periodic CB and W, are below.

\section{Convex 2D particles}
Let us now consider more general 2D-nanoparticles such as ellipses, or
disks which may have fused partly with the substrate, or any smooth
convex solid with a vertical axis of symmetry. Denote $D$ their
diameter (extension) parallel to the substrate plane, and $c>0$ the
random distance between neighboring nanoparticles. The nanoparticles
are identical, and the quenched distances $c_1,c_2,c_3,\dots$ make a
stationary sequence, in particular all $c_i$ have the same probability
density. Denote $F_{SL}(c_1,\dots,c_n)$ the free energy for such a
sequence, where on each interval the fluid chooses the configuration
of minimum energy, Cassie-Baxter or Wenzel. As for disks
or squares, there is a $c_{12}$ depending upon $\theta_0$ and upon the
shape of the nanoparticles, such that the Cassie-Baxter state is
chosen when $c<c_{12}$ and the Wenzel state is chosen when $c>c_{12}$. 
Denote $F_{SL}^{CB}(c_1,\dots,c_n)$ the free energy for the sequence
$c_1,\dots,c_n$, where on each interval the fluid chooses the
Cassie-Baxter configuration, and similarly
$F_{SL}^W(c_1,\dots,c_n)$ for the Wenzel case. Denote $1_{c<c_{12}}$ the indicator
function, with value 1 when the event is true and 0 otherwise. Then 
\begin{eqnarray*}
{1\over n}\langle F_{SL}(c_1\dots c_n)-F_{SL}^{CB}(c_1\dots c_n)\rangle
= \hskip2cm\cr
\langle 1_{c>c_{12}}(F_{SL}^W(c)-F_{SL}^{CB}(c))\rangle<0 \cr
{1\over n}\langle F_{SL}(c_1\dots c_n)-F_{SL}^W(c_1\dots c_n)\rangle
= \hskip2cm\cr
\langle 1_{c<c_{12}}(F_{SL}^{CB}(c)-F_{SL}^W(c))\rangle<0
\end{eqnarray*}
The solid-liquid free energy is less than the Cassie-Baxter or Wenzel
approximations. Therefore the contact angle $\theta$ will be less than
the approximations $\theta^{CB}$ or $\theta^W$. 
This demonstrates clearly and very generally that the disorder 
always reduces superhydrophobicity. 

The solution may be pursued further: $F_{SL}^W(c)$ is an affine
function of $c$ with slope $\gamma_{SL}$, 
and $F_{SL}^{CB}(c)$ is an affine function of  $c$ with slope
$\gamma_{SV}+ \gamma_{LV}$.  Using
$F_{SL}^{CB}(c_{12})=F_{SL}^W(c_{12})$ and Young's equation we get
\begin{equation} 
F_{SL}^W(c)-F_{SL}^{CB}(c)=-(1+\cos\theta_0 )(c-c_{12}). 
\end{equation} 
Therefore
\begin{equation} 
\langle 1_{c> c_{12}}(F_{SL}^W(c)-F_{SL}^{CB}(c))\rangle=
 -(1+\cos\theta_0 ) \langle 1_{c> c_{12}}(c-c_{12})\rangle
\end{equation} 
which leads to the same formulas (\ref{nanoCB}) as for disks, allowing
a computation of the effect of disorder on the contact angle.
Now $D$ has a more general definition, and $c_{12}$ may have to be
computed numerically. Like for disks, there is a $c_{\min}$ such that the Wenzel
state only exists when $c>c_{\min}$. One can check that $c_{12}>c_{\min}$
by the triangular inequality whatever the shape of the nanoparticles.
One can also prove in this generality that
$\cos\theta$ is a convex function of $\cos \theta_0$.
Indeed from Young's equation it is equivalent to show that
\begin{equation}
\int_0^{c_{12}}d\mu(c)F_{SL}^{CB}(c)+\int_{c_{12}}^\infty d\mu(c)F_{SL}^W(c)
\end{equation}
is a concave function of $\cos\theta_0$. This follows from
\begin{eqnarray*}
{1\over2\gamma_{LV}}{d^2F_{SL}^{CB}(c)\over d(\cos\theta_0)^2}=
-{R(\theta_0)\over\sin\theta_0}\hskip3.5cm\cr
{1\over2\gamma_{LV}}{d^2F_{SL}^W(c)\over d(\cos\theta_0)^2}=
-{4R(2\theta_0-\pi)\over\sin\theta_0}\hskip3cm\cr
-{1\over(\sin\theta_0)^3}\int_0^{2\theta_0-\pi}R(\theta)\sin\theta d\theta
\hskip1cm\end{eqnarray*}
where $R(\theta)=d\ell/d\theta$ is the radius of curvature at any point on the
curve defining the nanoparticle where the slope of the curve is $\tan\theta$.
Like for the circle on Fig.~\ref{fig4}, special points on the curve play 
a role: at a point in contact with the three phases, we have $\theta=\theta_0$
in the Cassie-Baxter case and $\theta=2\theta_0-\pi$ in the Wenzel case.

\section{Towards the 3D nanoparticles case}
\begin{figure}
\resizebox{0.78\textwidth}{!}{\includegraphics{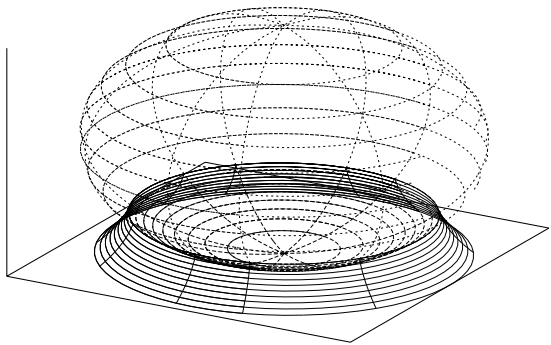}}
 \caption{\label{cat}Catenoid liquid-vapor interface}
 \end{figure}

The exact results presented so far are for 2D models or 3D grooves.
A true 3D situation will not be exactly soluble in general because
liquid penetration into one region is likely to influence
penetration into another region. 
There are however situations where the 3D nanoparticle case is soluble
and where the solution gives an insight into what may be specific to 2D
and what may extend to 3D. Consider once more Fig. \ref{fig4}, now
as a side view 
of a three-dimensional system. Suppose that $\theta_0$ and the random
distribution of nanoparticles on the plane surface are such that either
the Cassie-Baxter (left) configuration wins everywhere or the Wenzel (right)
configuration wins everywhere. Then the liquid-vapor interface either will be
plane everywhere, with spherical caps emerging, or will be made of catenoid
ribbons around each nanoparticle, as shown on Fig. \ref{cat}. Indeed
the catenoid is then the minimal  
surface, given the boundary conditions on the spheres. The resulting free 
energies and contact angles depend only on the density of nanoparticles.
Of course this is valid only if the variance of the disorder is sufficiently
small so that one or the other wins everywhere, a severe restriction
to the exact solution compared to the 2D case. 
 
\begin{figure}\begin{center}
\resizebox{0.35\textwidth}{!}{\includegraphics{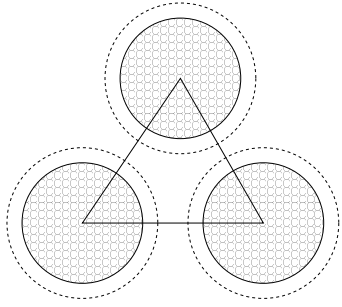}}
 \caption{\label{tri}Free energy decomposition in triangle}
 \end{center}\end{figure}
The arrangement of nanoparticles makes a random triangulation of the plane,
with the centers of the spheres as vertices of the triangulation.
The free energy per unit projected area for an extended system
is the same as the free energy per unit projected area for an equilateral
triangle corresponding to the given density of nanoparticles, 
or even for any triangle of same area, which is the average area of the 
triangles in the random triangulation, denoted $A_{\rm tri}$.
The free energies for a triangle in a dry, Cassie-Baxter, or Wenzel 
configuration are then computed exactly, using Fig. \ref{tri},
\begin{eqnarray}\label{F3D}
&F_{SV}^{rr}=(A_{\rm tri}+2\pi R^2)\gamma_{SV} \hskip4cm\cr\cr
&F_{SL}^{CB}=(A_{\rm tri}+\pi(1-\cos\theta_0)R^2)\gamma_{SV}+\hskip2.3cm\cr
&+\pi(1+\cos\theta_0)R^2\gamma_{SL}
+(A_{\rm tri}-{1\over2}\pi R^2\sin^2\theta_0)\gamma_{LV} \cr\cr
&F_{SL}^{W}=\Big({1\over2}\pi\rho_0^2
+\pi(R^2-R\sqrt{R^2-\rho_1^2})\Big)\gamma_{SV}+\hskip1.3cm\cr
&+\Big(A_{\rm tri}-{1\over2}\pi\rho_0^2+\pi(R^2+R\sqrt{R^2-\rho_1^2})\Big)
\gamma_{SL}+\cr
&+{1\over2}A_{\rm cat}\gamma_{LV}\hskip4cm
\end{eqnarray}
where $\rho_1$ and $\rho_0$ are the catenoid radii on the sphere
and the plane respectively, and $A_{\rm cat}$ is the area of the catenoid. 

\begin{figure}
\resizebox{0.78\textwidth}{!}{\includegraphics{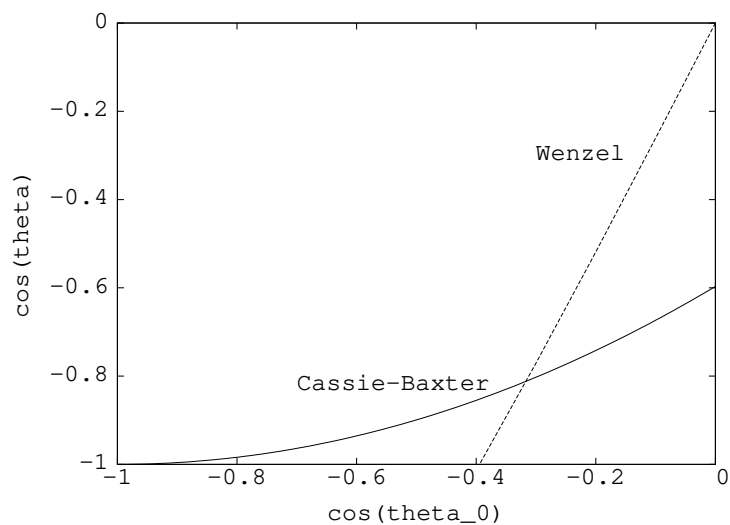}}
\caption{\label{costhe}3D-nanoparticles: $\cos\theta$ versus $\cos\theta_0$, 
for a density of nanoparticles equal to 0.5 divided by the area of 
an equilateral triangle of side $3R$.} 
\end{figure}

Let us now explain the derivation of (\ref{F3D}).
In the formula for $F_{SV}^{rr}$, the term $2\pi R^2$ corresponds to one half
of a sphere in contact with air, as is obvious from Fig.~\ref{tri}.
The formula for $F_{SL}^{CB}$ comes from
$$
F_{SL}^{CB}=(A_{\rm tri}+A_{\rm Vsph})\gamma_{SV}+A_{\rm Lsph}\gamma_{SL}
+A_{\rm LV}\gamma_{LV} 
$$
where $A_{\rm Vsph}=\pi(1-\cos\theta_0)R^2$ is the nanoparticle area in contact 
with the vapor in the Cassie-Baxter configuration, within the triangle; 
$A_{\rm Lsph}=\pi(1+\cos\theta_0)R^2$ is the nanoparticle area in contact 
with the liquid in the Cassie-Baxter configuration, within the triangle; 
$A_{\rm LV}=A_{\rm tri}-\pi\sin^2\theta_0R^2$ is the area of the liquid-vapor 
interface within the triangle in the Cassie-Baxter configuration.
The formula for $F_{SL}^W$ comes from
\begin{eqnarray*}
F_{SL}^W=(A_{\rm Vsph}+\12\pi\rho_0^2)\gamma_{SV}+\hskip3cm\cr
+\bigl(A_{\rm tri}+A_{\rm Lsph}-\12\pi\rho_0^2\bigr)\gamma_{SL}
+A_{\rm cat}\gamma_{LV} 
\end{eqnarray*}
where now $A_{\rm Vsph}$ is the nanoparticle area in contact 
with the vapor in the Wenzel configuration, within the triangle;
$A_{\rm Lsph}$ is the nanoparticle area in contact 
with the liquid in the Wenzel configuration, within the triangle, and
$A_{\rm cat}$ is the catenoid area within the triangle.
The equations of the sphere and of the catenoid, in cylindrical coordinates 
using the vertical axis of symmetry of the sphere, are respectively
$$
\rho=R\sqrt{1-\Bigl(1-{z\over R}\Bigr)^2},\quad 
\rho=\rho_c\cosh\Bigl({z-z_c\over\rho_c}\Bigr)
$$
where $z_c,\rho_c$ are free parameters. We denote $z_1,\rho_1$ the solution
of this system of equations, with smaller $z$, and we define $\alpha_1$ 
by $\cos\alpha_1=1-z_1/R$. The catenoid also intersects
the substrate plane at $z=0,\,\rho=\rho_0$. At $z_1,\rho_1$ and at 
$z=0,\,\rho=\rho_0$, the liquid contact angle must equal $\theta_0$,
solution of Young's equation 
$\gamma_{SV}=\gamma_{SL}+\gamma_{LV}\cos\theta_0$. This fixes 
the parameters $z_c,\rho_c$ and can be used to get numerical values for
$z_1,\rho_1,\alpha_1,\rho_0$. Then
\begin{eqnarray*}
A_{\rm Vsph}&=&\pi(1-\cos\alpha_1)R^2\cr
A_{\rm Lsph}&=&\pi(1+\cos\alpha_1)R^2\cr
A_{\rm cat}&=&\12\pi\rho_c^2\Bigl({z_1\over\rho_c}+\12\sinh{2z_c\over\rho_c}
+\12\sinh{2z_1-2z_c\over\rho_c}\Bigr)
\end{eqnarray*}
From (\ref{F3D}) and Young's equation 
$\gamma_{SV}^{rr}=\gamma_{SL}^{CB}+\gamma_{LV}\cos\theta^{CB}$ or
$\gamma_{SV}^{rr}=\gamma_{SL}^W+\gamma_{LV}\cos\theta^W$
the resulting contact angles $\theta^{CB}$ or $\theta^W$ can be computed 
exactly, and are shown on Fig. \ref {costhe}. The true contact angle $\theta$
on the random substrate will be $\min\{\theta^{CB},\theta^W\}$, or the maximum
of their cosines shown on the figure.

\begin{figure}
\resizebox{0.78\textwidth}{!}{\includegraphics{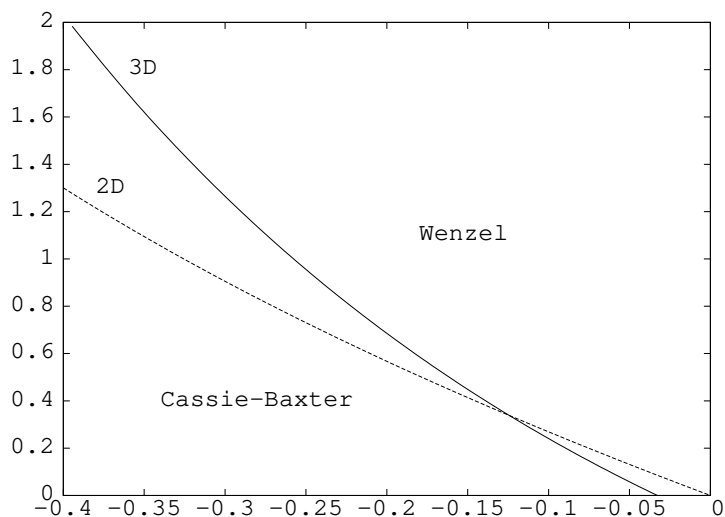}}
\caption{\label{tri12}Normalized excess area versus $\cos\theta_0$}
\end{figure}

We may also ask for conditions on a given triangle for a Cassie-Baxter or Wenzel
configuration to be a minimizer. The formulas above show that the area alone
can tell which of the two is the minimizer, as function of $\theta_0$.
The exact answer is shown on Fig. \ref{tri12}, where the 2D case has been
added for comparison.
The normalized excess area is the area of the triangle minus the area of
the close packing equilateral triangle, divided by $\pi R^2/2$ which is the 
projected area of the spheres on the plane inside the triangle. In 2D,
the normalized excess  length is the distance $c$ between disks
divided by the length $2R$ of the projection of a disk, with
coexistence at $c=c_{12}$. Fig. \ref{tri12} is to be used as follows:
given $\cos\theta_0$ and the normalized excess area or length of the
substrate with nanoparticles, 
if the point of corresponding abscissa and ordinate lies below the 3D
or 2D curve, then the stable state is the Cassie-Baxter state, otherwise it
is the Wenzel state. In Fig. \ref{costhe}, the normalized excess area
is $5\sqrt{3}/(2\pi)\simeq1.38$.

In 3D, a super-hydrophobic surface is to be expected when essentially all 
triangles have a normalized excess area below the coexistence curve,
because Cassie-Baxter states trap air more than Wenzel states,
but not far from the coexistence curve, because the excess area should be 
maximized in order to maximize the amount of air trapped. 
As Fig. \ref{tri12} shows, with a zoom around $\cos\theta_0=-0.05$,
a Cassie-Baxter state can be achieved whenever $\cos\theta_0<-0.034$ or 
$\theta_0>92^\circ$. 
But $\theta_0$ should be a little larger to leave some room 
for disorder and to achieve a larger final contact angle $\theta$.
In 2D, $\theta_0>90^\circ$ suffices, and the $d$-geometric model with $2d<c_{12}$
and $\langle c\rangle\simeq c_{12}$ achieves the desired goal.

\section{Conclusion}
We have thus generalized to different types of geometries the
classical formulas describing the Cassie-Baxter and Wenzel states for
wetting of rough substrates for 2D systems or 3D grooves geometries or 
3D-nanoparticles. Those include explicitly rectangular protrusions and
nanoparticles (or nanotubes). We have shown how to introduce the
concept of surface disorder within the problem. Our result, quite
generally, shows that disorder will lower  the contact angle observed
on such substrates compared to the periodic case.  We have also shown
that there are some choices of disorder for which the loss of
superhydrophobicity can be made relatively small, making
superhydrophobicity relatively robust.  
As an indirect result, this work may open the way for
new numerical simulation studies.

\section*{Acknowledgments}
The present work was done when 
F. Dunlop and T. Huillet were visiting professors at the University of Mons,
and when J. De Coninck was visiting professor at the University of 
Cergy-Pontoise. This work is partially supported by the Minist\`ere de
la R\'egion Wallonne and the Belgian Funds for Scientific Research (FNRS).

\end{document}